\documentclass[9pt, journal, twocolumn, twoside]{IEEEtran}

\usepackage[utf8]{inputenc}
\usepackage{helvet}
\usepackage{amsfonts}
\usepackage{amsmath}
\usepackage{amssymb}
\usepackage{cleveref}
\crefname{equation}{eq.}{eqs.}
\crefname{figure}{fig.}{figs.}
\crefname{table}{table}{tables}

\usepackage{graphicx}
\usepackage{sidecap}

\newcommand{\sswidth}{\columnwidth}
\newcommand{\swidth}{\columnwidth}
\newcommand{\mwidth}{\columnwidth}

\newcommand{\T}{\rule{0pt}{2.6ex}}
\newcommand{\B}{\rule[-1.0ex]{0pt}{0pt}}

\usepackage{threeparttable}
\usepackage{multirow}
\usepackage{setspace}
\usepackage{booktabs} 

\usepackage{color}
\usepackage{cite}

\begin{document}
\title{Statistically Segregated k-Space Sampling for \\ Accelerating Multiple-Acquisition MRI}
\author{L Kerem Senel, Toygan Kilic, Alper Gungor, Emre Kopanoglu, H Emre Guven, Emine U Saritas, Aykut Koc, Tolga~\c{C}ukur$^\ast$,~\IEEEmembership{Member,~IEEE}%
\thanks{This work was supported in part by a Marie Curie Actions Career Integration Grant (PCIG13-GA-2013-618101), by a European Molecular Biology Organization Installation Grant (IG 3028), by a TUBA GEBIP fellowship, and by BAGEP 2016 and BAGEP 2017 awards of the Science Academy. \textit{Asterisk indicates the corresponding author (e-mail: cukur@ee.bilkent.edu.tr).}}%
\thanks{E.U.Saritas and $^\ast$T.\c{C}ukur are with the Department of Electrical and Electronics Engineering, the National Magnetic Resonance Research Center (UMRAM), and the Neuroscience Program at Sabuncu Brain Research Center, Bilkent University, Bilkent, Ankara, Turkey. L.K.Senel and T.Kilic are with the Department of Electrical and Electronics Engineering and the National Magnetic Resonance Research Center (UMRAM), Bilkent University, Bilkent, Ankara, Turkey. L.K.Senel, A.Gungor, E.Kopanoglu, A.Koc and H.E.Guven are with the ASELSAN Research Center, Ankara, Turkey. E.Kopanoglu is with Brain Research Imaging Centre, Cardiff University, Cardiff, UK.}%
}
\maketitle

\begin{abstract}
A central limitation of multiple-acquisition magnetic resonance imaging (MRI) is the degradation in scan efficiency as the number of distinct datasets grows. Sparse recovery techniques can alleviate this limitation via randomly undersampled acquisitions. A frequent sampling strategy is to prescribe for each acquisition a different random pattern drawn from a common sampling density. However, naive random patterns often contain gaps or clusters across the acquisition dimension that in turn can degrade reconstruction quality or reduce scan efficiency. To address this problem, a statistically-segregated sampling method is proposed for multiple-acquisition MRI. This method generates multiple patterns sequentially, while adaptively modifying the sampling density to minimize k-space overlap across patterns. As a result, it improves incoherence across acquisitions while still maintaining similar sampling density across the radial dimension of k-space. Comprehensive simulations and in vivo results are presented for phase-cycled balanced steady-state free precession and multi-echo T$_2$-weighted imaging. Segregated sampling achieves significantly improved quality in both Fourier and compressed-sensing reconstructions of multiple-acquisition datasets. 
 
\end{abstract}


\section{Introduction}
\IEEEPARstart{M}{ultiple}-acquisition MRI methods are used when the image quality or information content of a single acquisition is insufficient. These methods acquire multiple images of the same anatomy, typically with different sequence parameters and image contrasts. Examples include phase-cycled balanced steady-state free precession (bSSFP) and multi-echo T$_2$-weighted imaging. Typical uses of multiple acquisitions include improved suppression of background tissues \cite{Koktzoglou:2009hp,Bangerter:2011ed}, relaxometry \cite{Deoni:2003dk}, extended slice coverage \cite{Saritas:2013bs}, separation of distinct resonances \cite{Reeder:2005gu,Cukur:2011iy}, and reduction of image artifacts \cite{Bangerter:2004hq,Cukur:2008ht}. While performance scales with the number of datasets acquired (N), this results in longer scan times and increased motion sensitivity. Therefore, multiple-acquisition methods can greatly benefit from acceleration techniques that enable high scan efficiency.   

Leveraging the sparse nature of MR images, compressed sensing (CS) techniques \cite{Lustig:2007cu,Gamper:2008da,Uecker:2008il,Jung:2009ir} were recently proposed to accelerate multiple-acquisition MRI. This powerful approach was demonstrated in several applications including fat-water separation \cite{Hernando:2010bz,Doneva:2010it,Wiens:2014fz}, parametric mapping \cite{Doneva:2010fe,Zhao:2015ez,Velikina:2013jb,Huang:2012hu,Sumpf:2011fn}, diffusion-weighting imaging \cite{Menzel:2011hm,Landman:2012ek,Haldar:2013fs}, subtraction angiography \cite{Rapacchi:2014ka}, multi-contrast imaging \cite{Bilgic:2011jv,Majumdar:2011hj,Huang:2014ca}, and lately bSSFP imaging \cite{Cukur:2015ic}. Individual acquisitions were accelerated via variable-density sampling patterns because the energy spectrum of MRI images is heavily constrained to central k-space \cite{Lustig:2007cu,Puy:2011gp}. Unacquired k-space data were then recovered via nonlinear reconstructions that enforce compressibility in a transform domain \cite{Knoll:2011hp,Murphy:2012hq,Liang:2009cd}.

The success of CS reconstructions depends critically on the selection of k-space sampling locations. Much work has been done on optimizing sampling patterns for single-acquisition MRI. Theory indicates that random patterns that promote incoherent aliasing guarantee sparse recovery with high probability \cite{Donoho:2006ja,Candes:2007es}. Thus, many early studies proposed variable-density random patterns to maximize incoherence of aliasing artifacts in spatial or temporal dimensions \cite{Lustig:2007cu,Jung:2009ir,Hu:2010bb,Cukur:2011kr}. Improved strategies were later considered to maintain a favorable compromise between incoherence and practical imaging considerations. For instance, pattern formation based on adaptive density estimation was suggested to effectively utilize prior information about the energy spectrum of specific datasets \cite{Knoll:2011be,Liang:2012cq,Raja:2014gg}. Recent studies also imposed deterministic constraints on sampling patterns to prevent unwanted gaps or clusters. Examples of this approach include Poisson disc sampling to improve uniformity of inter-sample distances in multi-coil imaging \cite{Lustig:2010hs}, optimization routines to maintain fixed frame rates in dynamic imaging \cite{Ahmad:2015iz, Kim:2015fu}, complementary Poisson sampling for variable view sharing in dynamic contrast-enhanced imaging \cite{Levine:2016jt}, and sample ordering to minimize eddy-current artifacts in segmented acquisitions \cite{Tamir:2016eq}. Hybrid strategies were also proposed that deploy deterministic sampling in central and random sampling in peripheral regions of k-space to better suppress aliasing artifacts in reconstructed images \cite{Sung:2013co}. 

Contrary to single-acquisition MRI, less attention has been given to sampling patterns for multiple static acquisitions. A frequent strategy is to accelerate each acquisition via a different random pattern drawn from a common sampling density \cite{Doneva:2010it,Bilgic:2011jv,Huang:2014ca}. Because separate instances of a random variable are independent, this strategy is expected to yield incoherent aliasing across acquisitions \cite{Lustig:2007cu}. Yet, naive random selection often yields gaps or clusters in the acquisition dimension that can degrade reconstruction quality or reduce scan efficiency. In a recent study on phase-cycled bSSFP imaging, we proposed a low-correlation sampling method to limit gap or cluster formation across acquisitions \cite{Cukur:2015ic}. Candidate sets of patterns were first generated, and the set with the lowest inter-pattern correlations was searched. This method achieved a modest decrease in pattern overlap \cite{Cukur:2015ic}, but a heuristic search among uninformed patterns is computationally intensive and suboptimal. 

Here we aim to minimize k-space overlap across patterns, while maintaining similar sampling density across multiple acquisitions. To achieve this goal, we devise a mathematical framework for statistically-segregated sampling in multiple-acquisition MRI. The proposed method generates each of N patterns in sequence, while adaptively modifying the sampling density to promote minimal pattern overlap. For each pattern, the sampling probability is lowered for k-space locations that are covered by preceding patterns. The probability for uncovered locations is appropriately increased to maintain identical sampling density across the radial k-space dimension within each pattern. Segregated sampling preserves the stochastic nature of individual patterns while effectively increasing k-space coverage. It significantly reduces pattern overlap compared to random sampling, without the need for time-consuming search or optimization procedures. Simulations and in vivo results on phase-cycled bSSFP and multi-contrast imaging clearly demonstrate improved quality in Fourier and CS reconstructions of multiple-acquisition datasets.

\section{Theory}
\subsection*{Multiple-Acquisition MRI}
Here we consider two multiple-acquisition applications, phase-cycled bSSFP imaging and multi-contrast T$_2$-weighted imaging. Main field inhomogeneities can introduce regions of signal voids in bSSFP images known as banding artifacts \cite{Bangerter04}. To prevent signal loss, phase-cycled bSSFP methods acquire multiple images of the same anatomy with nearly identical contrast except for a spatial shift in the location of artifacts \cite{WCSSFP}. This is implemented by applying a unique phase-cycling value between consecutive RF pulses during each acquisition. The resulting bSSFP signal can be expressed as,
\begin{equation}
\label{eq:ssfp}
S_{n}(\vec{r}) = {M(\vec{r})}\frac{{{e^{i\left( {\phi(\vec{r})  + \Delta {\phi _n}} \right)/2}} \left( {1 - A(\vec{r}){e^{ - i\left( {\phi(\vec{r})  + \Delta {\phi _n}} \right)}}} \right)}}{{1 - B(\vec{r})\cos (\phi(\vec{r})  + \Delta {\phi _n})}}
\end{equation}
under the assumption that the echo time (TE) is half the repetition time (TR). In 
\Cref{eq:ssfp}, $\vec{r}$ denotes the spatial location vector, $\phi$ is the phase accrued in a TR due to field inhomogeneity, and $\Delta \phi_n$ is the phase-cycling value selected for the n$^{th}$ acquisition (n $\in$ [1 N]). $M$, $A$, $B$ that do not depend on field inhomogeneity are described elsewhere \cite{Cukur:2015ic}. Multiple bSSFP acquisitions with differential sensitivity to field inhomogeneity carry similar information about tissue structure. These acquisitions can be simply combined \cite{PNORM, Quist:2012kx} or jointly reconstructed \cite{Ilicak:FJpKoYYb} to suppress banding artifacts.  

Spin-echo (SE) imaging with T$_2$-weighted contrast is pervasive in anatomical assessment. However, a single T$_2$-weighting may be suboptimal when relaxation parameters vary substantially across subjects \cite{Gold:2004ij} or when tissues show relatively broad variation in T$_2$ values. In such cases, multiple T$_2$-weighted images with varunyg TE values can be colllected. The resulting SE signal is \cite{Handbook},
\begin{equation}
\label{eq:se}
S_{n}(\vec{r}) = {iM(\vec{r})} \cdot (1-e^{-{TR}/{T_1(\vec{r})}}) \cdot (e^{-{TE _n}/{T_2(\vec{r})}})
\end{equation}
under the assumption that TR $\gg$ TE$_n$. In the above equation, T$_1(\vec{r})$ and T$_2(\vec{r})$ denote the spatial distribution of longitudinal recovery and transverse relaxation time constants, respectively. TE$_n$ denotes the echo-time of the respective SE acquisition where n $\in$ [1 N]. Because multiple SE acquisitions with differential T$_2$-weighting as in \Cref{eq:se} carry shared tissue information (e.g., location of tissue boundaries), they can be jointly reconstructed \cite{Majumdar:2011hj,Bilgic:2011jv} to improve image quality and to enhance tissue discrimination \cite{Misaki:2015iy}. 

Prescribing a larger N significantly improves image quality in both bSSFP and multi-contrast applications. Meanwhile, undesirable lengthening of scan times can be prevented through k-space undersampling. The unacquired data can then be estimated by solving an inverse problem based on the following forward model: 
\begin{equation}
\label{eq:invprob}
{y_n}(\vec{k}) = D_n \mathcal{F} \left\{ { {S_n}(\vec{r})} \right\}
\end{equation}  
Here $y_n$ denotes the k-space data for the n$^{th}$ acquisition, $\vec{k}$ is the k-space location vector, $\mathcal{F}$ is the Fourier-transformation, and $D_n$ is a binary mask that reflects the $n^{th}$ sampling pattern. 

\subsection*{Variable-Density Random Sampling}
The energy spectrum of MRI images follow an approximate power-law in k-space \cite{Puy:2011gp}. The transform domain coefficients also tend to be sparser at fine-scales that reflect high spatial frequencies \cite{Lustig:2007cu}. As a result, variable-density random sampling (VDS) has come forth as a preferred companion to CS reconstructions. In VDS, the expected sampling density function (PDF) is specified to maintain a desired acceleration rate. For multiple-acquisition MRI, the sampling PDF is usually taken to be identical across acquisitions:
\begin{equation}
{p_{D{}_n}}({k_y},{k_z}) = {p_o}({k_r})
\end{equation}
where ${k_r} = \sqrt {k_y^2 + k_z^2}$ is the k-space radius, $p_o$ is the common density, and circular symmetry is assumed across phase-encoding dimensions without loss of generality. This density is then used to draw random instances of sampling patterns for each acquisition, i.e.,
\begin{equation}
{p_{D{}_n}}({k_y},{k_z})\xrightarrow{{draw}}{D_n}({k_y},{k_z})
\end{equation}

Although random sampling often yields a high degree of incoherence, the generated patterns may occasionally have poor aliasing properties. As a remedy, Monte-Carlo designs have been proposed where multiple sets of candidate patterns are drawn, $D_n^c({k_y},{k_z})$ \cite{Lustig:2007cu}. The incoherence of each pattern is measured via its point spread function (PSF). An image containing a unit-intensity voxel is undersampled in k-space with the given pattern, and re-transformed to the image domain to calculate the PSF. The ratio of peak intensity to maximum side-lobe intensity of the PSF ($R_{PSF}$) reflects incoherence. In this random sampling method, the candidate pattern with the maximum $R_{PSF}$ is selected for each acquisition independently:
\begin{equation}
 D_n =  {\mathop {\max }\limits_{D_n^c}{R }_{PSF}(D_n^c) } 
\end{equation}

Other desired properties can also be enforced, such as minimal correlation among patterns for multiple acquisitions \cite{Cukur:2015ic}. In this low-correlation method, a large number of candidates are first generated for the set of multiple patterns across acquisitions. The set of patterns with minimum pair-wise correlations can then be identified through brute-force search: 
\begin{equation}
 \{D_1,..,D_N\} = {\mathop {\min }\limits_{D_{1,..,N}^c} \left( \sum\limits_{i = 1}^N {\sum\limits_{j = i + 1}^N {corr(D_i^c,D_j^c)}}  \right)} 
\end{equation}

\begin{figure}[t]
  \begin{center}
    \includegraphics[width = \swidth]{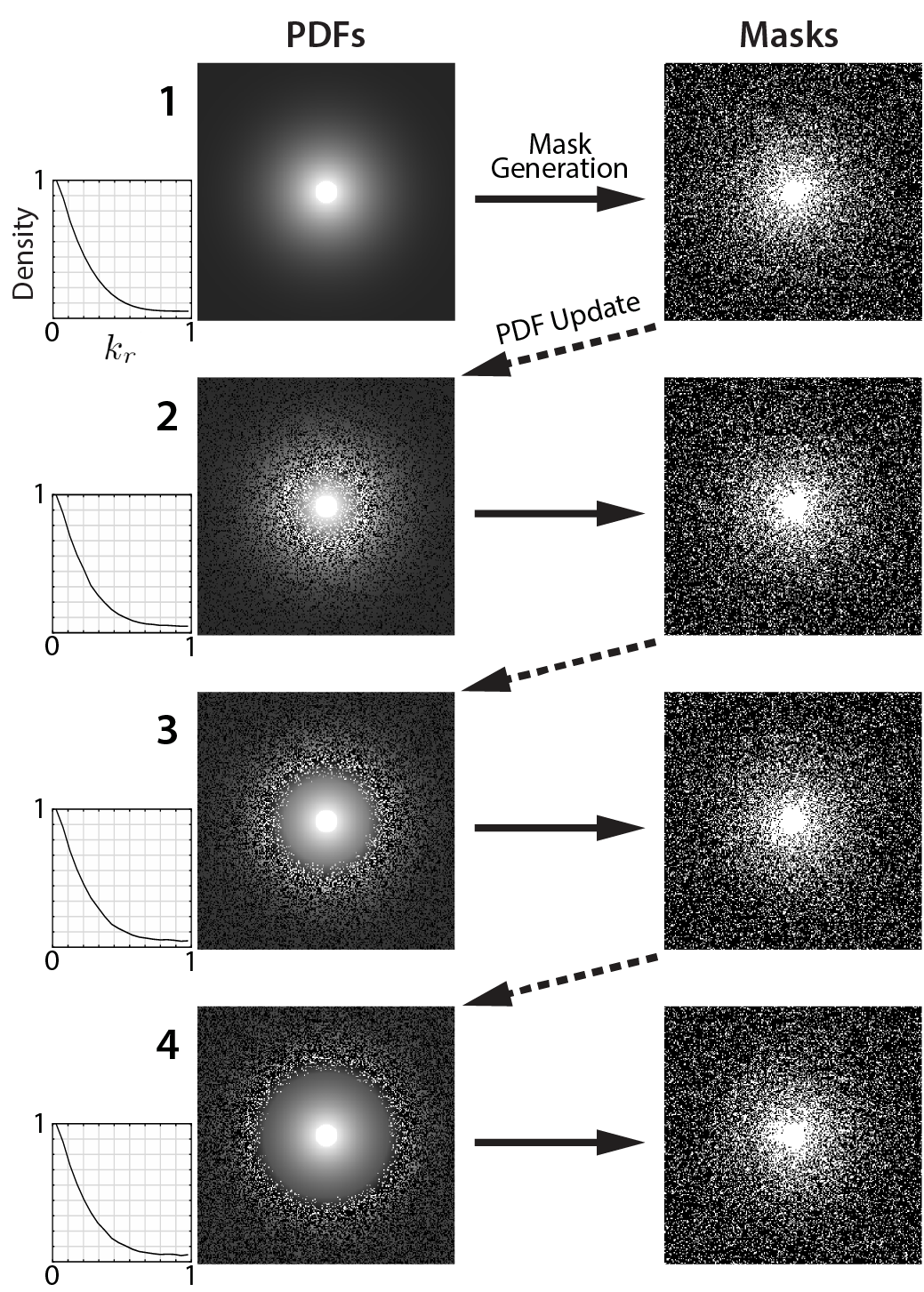}
    \caption{Segregated sampling designs N random undersampling patterns (Masks) via Monte-Carlo simulations based on respective sampling density functions (PDFs). Unlike standard random sampling, it adaptively modifies the sampling density to increase aggregate k-space coverage and to promote minimal pattern overlap. It lowers sampling density in k-space locations that are readily covered in preceding patterns, and increases sampling density for uncovered locations. At a given k-space radius ($k_r$), the total increase in density for uncovered locations is equal to the total decrease for covered locations. This procedure yields incoherent patterns across the acquisition dimension while maintaining identical sampling density across $k_r$.       
    }
    \label{fig:framework}
  \end{center}
\end{figure}

\subsection*{Sampling Performance: Coverage and Overlap}
Given their stochastic nature, random patterns are best analyzed in a statistical framework. In random sampling, patterns are generated independently and the inclusion of a k-space location within each pattern follows a Bernoulli distribution (with parameter $p_o(k_r)$). Thus, the total number of times (t) a k-space location is sampled across N acquisitions follows a binomial distribution:
\begin{equation}
{P_t (k_y, k_z)} = \left( {\begin{array}{*{20}{c}}
  N \\ 
  t 
\end{array}} \right){p_o}{{({k_r})}^t}{{\left[ {1 - {p_o}({k_r})} \right]}^{(N - t)}}
\end{equation}
The probability of complete omission from all acquisitions is:
\begin{equation}
{{P_0 (k_y, k_z)} = {{\left( {1 - {p_o}({k_r})} \right)}^N}}
\end{equation}
The probability of being sampled within a single acquisition is:
\begin{equation}
{{P_1 (k_y, k_z)} = N {{\left( {1 - {p_o}({k_r})} \right)}^{N-1}} {p_o}({k_r})}
\end{equation}

Here we give attention to three properties of multiple-acquisition patterns: aggregate coverage, differential coverage, and overlap. We take the aggregate coverage of N patterns as the proportion of k-space locations that are sampled in at least one pattern: 
\begin{equation}
\label{eq:prop1}
\%\mbox{ coverage}  = \frac{1}{T}\sum\limits_{{k_y},{k_z}} {(1 - P_0 (k_y, k_z))} 
\end{equation}
where $T$ denotes the total number locations in the sampling grid. Meanwhile, we take differential coverage as the proportion of locations that are uniquely sampled within an individual pattern:
\begin{equation}
\label{eq:prop2}
\%\mbox{ differential cov.} = \frac{1}{T}\sum\limits_{{k_y},{k_z}} {(P_1 (k_y, k_z))} 
\end{equation}
Lastly, we take overlap among patterns as the number of times a location has been repeatedly sampled across acquisitions:
\begin{equation}
O = \left\{ {\begin{array}{*{20}{c}}
  {t - 1,\mbox{ if } t \geqslant 2} \\ 
  {0,\mbox{ otherwise}} 
\end{array}} \right.
\end{equation}
The percentage overlap is then measured as the expected value of $O$ averaged across k-space:
\begin{equation}
\label{eq:prop3}
\%\mbox{ overlap} = \frac{1}{{T.(N - 1)}}\sum\limits_{{k_y},{k_z}} {\sum\limits_{t = 2}^N {(t - 1).{P_t (k_y, k_z)}} } 
\end{equation}

Eq.~\ref{eq:prop1} clearly shows that aggregate coverage decreases towards high spatial frequencies (i.e., lower sampling density) and with smaller N. If aggregate coverage is broadened by increasing either the sampling density or N, the range of achievable acceleration factors will be limited and the total scan time will be prolonged. Note that higher sampling density and larger N will result in decreased differential coverage (Eq.~\ref{eq:prop2}) and increased pattern overlap (Eq.~\ref{eq:prop3}). As a result, redundant or highly similar information will be collected across acquisitions, reducing scan efficiency. This inherent trade-off poses a significant limitation on the utility of random sampling.

\subsection*{Statistically Segregated k-space Sampling}
Here we propose a statistically-segregated sampling method that broadens the aggregate coverage of multiple patterns to increase the amount of tissue information captured. In random sampling, major increases in $p_o(k_r)$ or N are required to boost coverage, but these changes in turn prolong scan times. In segregated sampling, coverage is instead enhanced by controlling for unwanted overlap across patterns while retaining the same N and similar radial sampling density in each pattern. 

The proposed method is implemented via a statistical framework (Fig.~\ref{fig:framework}) where the joint probability distribution for N patterns is:
\begin{equation}
\begin{aligned}
&{p_{{D_1},..,{D_N}}}({k_{y1,z1}},..,{k_{yN,zN}}) = {p_{{D_N}|{\bar D_{N - 1}}}} \cdot .. \cdot {p_{{D_2}|{\bar D_1}}}\cdot{p_{{D_1}}} \\
&\mbox{subj. to } \quad {p_{D{}_n|{\bar D_{n - 1}}}}({k_r}) = \int\limits_{{k_\theta }} {{p_{{D_n}|{\bar D_{n - 1}}}}({k_r},{k_\theta })}  = {p_o}({k_r}) 
\end{aligned}
\end{equation}
where $\bar D_{n} = \{ D_n,..D_1\}$. The joint distribution is decomposed into conditional distributions, constrained to follow a pre-selected sampling density ($p_o$) across radial k-space ($k_r$). The conditional distributions and respective sampling patterns are generated sequentially, starting with $D_1$:
\begin{equation}
{p_{{D_1}}}({k_{y,z}}) = p_{{o}}({k_r})
\end{equation}
Overlap in subsequent patterns is minimized by decreasing the sampling density ($p^-$) in previously covered locations while increasing it ($p^+$) in uncovered locations: 
\begin{eqnarray}
\label{eq:dmod1}
    {p^{-}_{D_n | {\bar D_{n - 1}}}(k_{y,z}^-)} &=& p_o({k_r}) \cdot \mu \\
    {p^{+}_{D_n | {\bar D_{n - 1}}}(k_{y,z}^+)} &=& {p_o({k_r})} \cdot {\beta_{n-1}(k_{r})}
\end{eqnarray}
where $k_{y,z}^-$ denotes locations covered at least once in previous patterns (i.e., $\sum\limits_{i = 1}^{n-1} {{D_i} \geqslant 1 }$), and $k_{y,z}^+$ denotes the remaining uncovered locations. The density modification is controlled via the parameter $\mu \in [0 \mbox{  } 1]$, which results in random sampling at $\mu = 1$ and maximally segregated sampling at $\mu = 0$. Meanwhile, $\beta_{n}$ is selected as:
\begin{equation}
{\beta _n}({k_{r}}) = \frac{1 - \mu \cdot K_n^-(k_r)}{1 - K_n^-(k_r)}
\end{equation}
For a given $k_r$, $K_n^-$ denotes the ratio of the number of unique locations sampled in patterns $\{D_1,..D_n\}$ to the number of locations on the sampling grid. This $\beta_{n}$ definition ensures that a fixed radial sampling density -in accordance with $p_o(k_r)$- is maintained (see Fig.~\ref{fig:framework} for example).

The proportion of grid locations sampled, $K_n^-$, grows steadily with $n$, and the growth rate depends on $p_o$ and varies across $k_r$. This rate can be examined by calculating its expected value $e_n (k_r)= E \left\{ K_n^-  (k_r)\right\} $. As expected $e_1 = p_o$, and for subsequent acquisitions:
\begin{equation}
\label{eq:diff}
\begin{aligned}
    {e_{n}} &= {e_{n-1}}+{(1-e_{n-1})}{E \left\{ p^{+}_{D_n | {\bar D_{n - 1}}} | K_{n-1}^- \right\} } \\
               &= {e_{n - 1}} + (1 - {e_{n - 1}}).\left( {{p_o}\frac{{1 - \mu {e_{n - 1}}}}{{1 - {e_{n - 1}}}}} \right) \\
               &= {e_{n-1}}{(1-\mu p_o)}+ p_o
\end{aligned}
\end{equation}
The solution of the difference equation in Eq.~\ref{eq:diff} is:
\begin{equation}
    e_{n} = \frac{1}{\mu} - \frac{1}{\mu} { ( 1 - \mu p_o )}^n
\end{equation}
Based on the expression above, it is possible to have $e_{n-1} < 1$ and $e_n \geq 1$ for a finite value of $n$ (equivalently $K^-_{n-1} < 1$ and $K^-_n = 1$). However, $e_n \geq 1$ suggests that the modified probability values $p^{+}_{D_n | {\bar D_{n - 1}}}$ exceed 1. When this violation is detected, a corrected rule is used for density modification instead of Eq.~\ref{eq:dmod1}:
\begin{eqnarray}
    {p^{-}_{D_n | {\bar D_{n - 1}}}(k_{y,z}^-)} &=& \frac{K_n^-(k_r) -1 + p_o({k_r}) }{K_n^-(k_r)} \\
    {p^{+}_{D_n | {\bar D_{n - 1}}}(k_{y,z}^+)} &=& 1
\end{eqnarray}
This updated rule ensures that the maximum possible density value is 1, and the density for the remaining locations is adjusted to maintain $p_o(k_r)$ in the radial dimension. After a certain number of acquisitions, no further segregation will be possible since $K_n^-(k_r) = 1$ for a finite value of $n$. At that point, all k-space locations at $k_r$ are assigned the sampling density $p^{-} = p_o(k_r)$. Thus subsequent patterns are drawn from the original sampling density.  

\begin{figure}[t]
 \begin{center}
    \includegraphics[width = \mwidth]{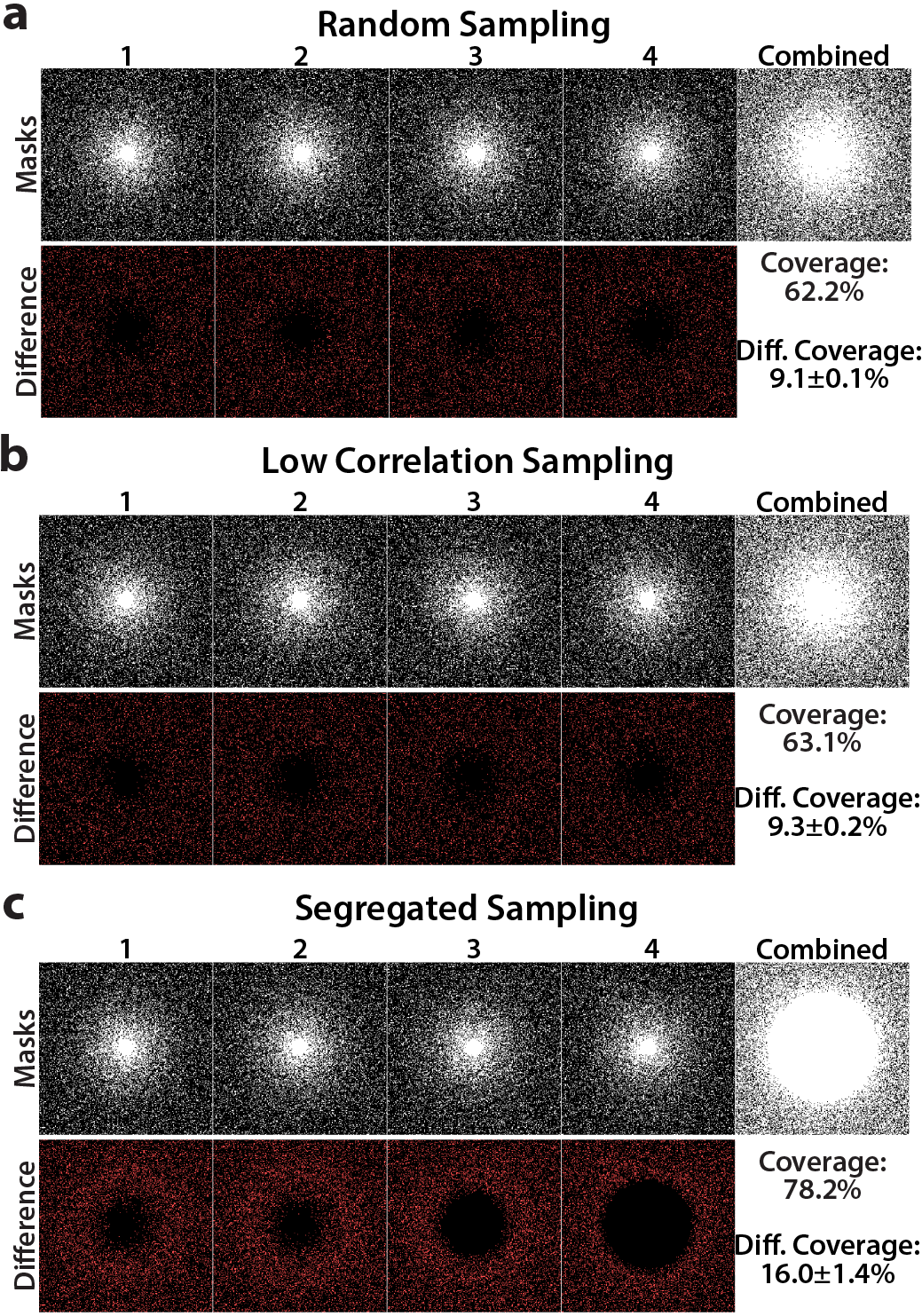}
    \caption{Representative sampling patterns for N=4, R=4 generated using \textbf{(a)} random sampling, \textbf{(b)} low-correlation sampling, and \textbf{(c)} segregated sampling. The resulting patterns and the aggregate pattern (Combined) are shown in upper rows. The difference masks comprising locations that are uniquely sampled by each pattern are shown in bottom rows. Segregated sampling increases aggregate k-space coverage to 78.2\% from merely 62.4\% in random and 63.1\% in low-correlation sampling. It also increases the average differential coverage 16.0$\pm$1.5\% (mean$\pm$std across N) from 9.1$\pm$0.1\% in random and 9.3$\pm$0.2\% in low-correlation sampling, due to reduced pattern overlap.
    }
    \label{fig:masks}
    \end{center}
\end{figure}

\section{Methods}
\subsection*{Generation of Sampling Patterns}
Random, low-correlation, and segregated sampling patterns were generated. A common PDF was designed to achieve a target acceleration factor (R) isotropically in the two phase-encode dimensions. The PDF was designed based on a polynomial function of k-space radius \cite{MikiCS}; the polynomial degree monotonically increased with R: degrees were (2, 3, 4, 5, 6) for R = (2, 3, 4, 6, 8). Depending on R, a central k-space region reaching 4\% to 18\% of the maximum spatial frequency was sampled at the Nyquist rate. For the brain phantom, the sampling grid sizes were 434$\times$362 in T$_1$-weighted and bSSFP acquisitions, and 362$\times$434 in T$_2$-weighted acquisitions. For in vivo experiments, the grid sizes were 256$\times$256 for bSSFP acquisitions and 192$\times$224 for T$_2$-weighted acquisitions.

Patterns were drawn from the designed PDFs via a Monte Carlo procedure described previously \cite{Cukur:2011kr}. For random and segregated sampling, the pattern that minimized aliasing energy was selected among 1000 candidate instances. For low-correlation sampling, 500 minimal-aliasing candidates were first generated by simulating 10000 pattern instances. The set of N candidates that yielded minimal pair-wise correlations was then selected \cite{Cukur:2015ic}.

\subsection*{Reconstruction of Multiple-Acquisition Data}
Two different reconstruction were performed on undersampled data. First, Fourier reconstructions of individual acquisitions (ZF) were computed: unacquired data were filled with zeros, data were compensated for the variable sampling density across k-space, and lastly an inverse Fourier transformation was taken. 

Second, all acquisitions were reconstructed jointly via a profile-encoding (PE) method that was recently proposed for multiple-acquisition imaging \cite{Ilicak:FJpKoYYb}. Inspired by the iterative self-consistent parallel imaging (SPIRiT) reconstruction for coil arrays \cite{Lustig:2010hs}, PE aims to express a given sample in each acquisition as a weighted combination of neighboring samples across all acquisitions. For this purpose, an interpolation kernel ($\mathcal{K}_n$) is estimated from calibration data in the central region of k-space. This kernel then synthesizes unacquired data as a weighted linear combination of acquired data. PE was implemented through the following optimization problem: 
\begin{equation}
\label{eq:perecon}
\begin{aligned}
\mathop {{\text{min}}}\limits_{{m_{1,..,N}}} \quad & \sum\limits_n {\left\{ {\left\| {{y_n} - {\mathcal{F_P}_n}\left\{ {{m_n}} \right\}} \right\|_2^2 + {\lambda _0}\left\| {({\mathcal{G}_n} - I){m_n}} \right\|_2^2} \right\}}  \\
&+ {\lambda _1}{\left\| {\sqrt {\sum\limits_n {{{\left| {\psi \{ {m_n}\} } \right|}^2}} } } \right\|_1}\\
\end{aligned}
\end{equation}  
where $m_n$ is the reconstructed image for the $n^{th}$ acquisition, and $\mathcal{G}_n$ is the image-domain equivalent of $\mathcal{K}_n$. The first term in the objective enforces the consistency of acquired data ($y_n$) with the reconstructed data (${\mathcal{F_P}_n}\left\{ {{m_n}} \right\}$). The second term enforces the consistency of the interpolation kernel ($\mathcal{G}_n$) with the reconstructed images ($m_n$). The third term is used to enforce joint-sparsity of the reconstructed images in a known transform domain ($\psi$) \cite{Murphy:2012hq}.

Here, $\mathcal{K}_n$ was estimated for a 11$\times$11 k-space neighborhood. Although the fully-sampled k-space radius varied between 4\% to 18\%, a broader region is sampled approximately at the Nyquist rate for variable-density patterns. Thus to effectively use information in acquired data,  $\mathcal{K}_n$ was trained in a calibration region of size 96$\times$96. Tykhonov regularization with weight $\alpha=0.01$ was used during training. The operator $\psi$ was a Daubechies 4 wavelet. PE in Eq.~\ref{eq:perecon} was decomposed into two subproblems using variable splitting with a splitting parameter of 1. The first problem containing the data and calibration consistency terms was solved via a conjugate gradient (CG) algorithm \cite{Murphy:2012hq}. The second problem containing the sparsity term was solved via soft thresholding. A total of 30 outer iterations yielded stable results. For phantom data, $\lambda_0$=$10^{-6}$, $\lambda_1$=$0$, 1 inner CG iteration were used. For in vivo data, $\lambda_0$=$10^{-6}$, $\lambda_1$=$5$x$10^{-4}$, 10 CG iterations (bSSFP) and 1 CG iteration (multi-contrast) were used. Prior to reconstruction, data were normalized to set the norm of density-compensated data divided by the square root of N to 1.

\begin{figure}[t]
  \begin{center}
    \includegraphics[width = \swidth]{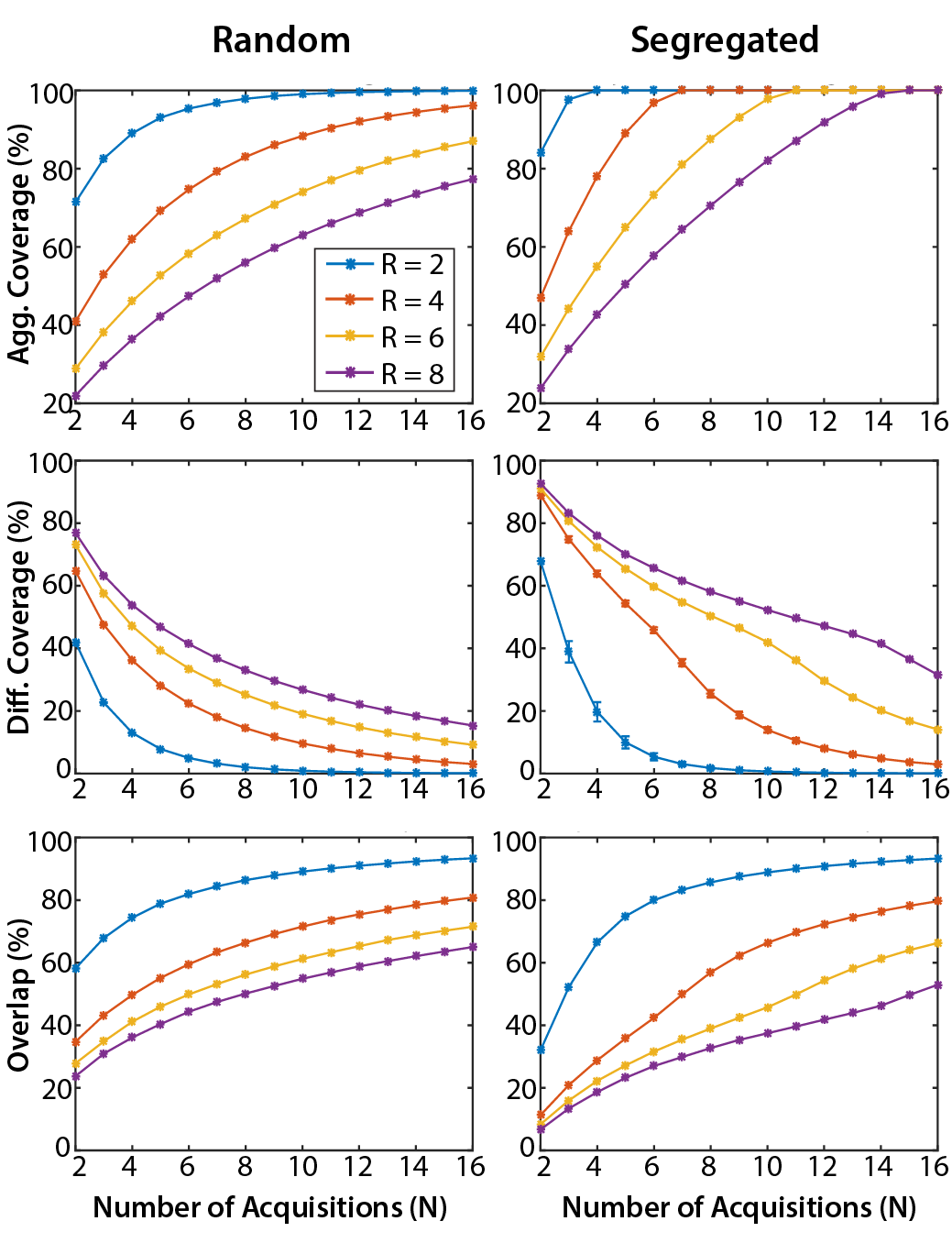}
    \caption{To assess sampling performance, patterns were generated via random (left column) and segregated sampling (right column) for N = [2 16] and R = [2 8]. Aggregate coverage: while random sampling leaves 10-30\% of k-space uncovered even at N = 16, segregated sampling achieves full coverage within N = 2$\times$R acquisitions. Differential coverage: compared to random sampling, segregated sampling significantly expands the portion of k-space uniquely covered by each pattern. Error bars show mean$\pm$std. of differential coverage across N patterns. Percentage overlap: segregated sampling yields reduced overlap, particularly for higher R and lower N.    
    }
    \label{fig:patcomp}
  \end{center}
\end{figure}

\subsection*{Simulations}
To theoretically assess sampling performance, sampling strategies were compared in terms of their aggregate coverage (\Cref{eq:prop1}), differential coverage (\Cref{eq:prop2}) and percentage overlap (\Cref{eq:prop3}). Random and segregated sampling patterns were generated for N = [2 16] and R = [2 8]. Differential coverage and overlap were normalized by the maximum coverage of a single pattern dictated by R. 
 
Sampling performance was assessed on a brain phantom at 0.5 mm isotropic resolution (http://www.bic.mni.mcgill.ca/brainweb). A single T$_1$-weighted image was simulated based on Eq.~\ref{eq:se}. Multiple acquisitions were obtained from this image by using N patterns, each with an undersampling factor of N. Random and segregated ($\mu = 0$) sampling were used to undersample in two phase-encode dimensions. The following T$_1$/T$_2$ values were used: 2570/330 ms for cerebro-spinal fluid (CSF), 1200/250 ms for blood, 500/70 ms for white matter, 830/83 ms for gray matter, 970/50 ms for muscle, and 350/70 ms for fat \cite{Stanisz:2005fe}. The parameters of SE acquisitions were $\alpha = 90^o–180^o$ (excitation and refocusing pulses), TR = 575 ms, and TE = 14 ms. ZF reconstructions were summed across acquisitions. 

Next, simulations were performed to demonstrate segregated sampling in the presence of variations in image structure across acquisitions. In phase-cycled bSSFP simulations, the signal for each tissue was calculated using Eq.~\ref{eq:ssfp}. The following T$_1$/T$_2$ values were used: 3000/1000 ms for cerebro-spinal fluid (CSF), 1200/250 ms for blood, 1000/80 ms for white matter, 1300/110 ms for gray matter, 1400/30 ms for muscle, and 370/130 ms for fat \cite{Gold:2004ij}. Meanwhile, the PD values were: 1 for CSF, blood, muscle and fat, 0.77 for white matter, and 0.86 for gray matter. The bSSFP sequence parameters were set to $\alpha = 45^o$ (flip angle), TR/TE = 5.0/2.5 ms, and $\Delta\phi $ spanning the range [0 2$\pi$) in steps of size $2\pi/N$. A main field inhomogeneity map was used that yielded off-resonance shifts of 0$\pm$62 Hz (mean$\pm$std across the volume). Balanced SSFP acquisitions were undersampled by a factor of N in two phase-encode dimensions using random, low-correlation, and segregated ($\mu = 0$) sampling. PE reconstructions were performed for N = 2, 4, 6, 8 and R = N. Individual phase-cycled images were p-norm combined across acquisitions (p = 2). 

\begin{table}[t]
	\scriptsize
	\setlength{\tabcolsep}{3.6pt}
	\centering
	\caption{Sampling Performance: Segregated versus Random}
	\label{tab:patcomp} 
	\begin{threeparttable}
	\begin{tabular*}{\columnwidth}{@{\extracolsep{\fill}}l|l|cccccc} 
		\multicolumn{8}{c}{} \\[0.25ex] 
\hline  
\multicolumn{2}{c|}{\T \textbf{}} & \textbf{N = 2} & \textbf{N = 3} & \textbf{N = 4} & \textbf{N = 6} & \textbf{N = 8} & \B \textbf{N = 10} \\
\hline
\hline \T
\multirow{3}{*}{\textbf{R=2}}   & Agg. & 12.7 & 15.1 & 10.9 & 4.5 & 2.0  & 1.0\\ 
                                & Diff. & 25.4 & 16.1 & 6.4 & 0.4 & -0.3 & -0.2 \\ 
                                & Over. & -25.4 & -15.1 & -7.1 & -2.0 & -0.7 & -0.2 \\ 
\hline \T
\multirow{3}{*}{\textbf{R=4}}   & Agg. & 5.9 & 11.3 & 15.9 & 22.3 & 17.1 & 11.7 \\ 
                                & Diff. & 23.2 & 27.1 & 27.2 & 23.4 & 10.8 & 4.2 \\ 
                                & Over. & -22.7 & -22.0 & -20.5 & -16.7 & -9.1 & -4.8 \\ 
\hline \T
\multirow{3}{*}{\textbf{R=6}}   & Agg. & 2.9 & 6.1 & 9.1 & 15.1 & 20.4 & 23.9\\ 
                                & Diff.  & 18.1 & 22.7 & 24.5 & 25.9 & 25.5 & 23.0\\ 
                                & Over. & -18.6 & -18.7 & -18.3 & -18.0 & -17.3 & -15.9 \\ 
\hline \T
\multirow{3}{*}{\textbf{R=8}}   & Agg. & 2.0 & 4.3 & 6.2 & 10.7 & 14.9 & 19.1\\ 
                                & Diff. & 16.5 & 21.1 & 22.6 & 24.3 & 25.4 & 25.5\\ 
                                & Over. & -16.7 & -17.1 & -16.7 & -16.3 & -16.7 & -16.6 \\ 
\hline                                            
\end{tabular*}
\begin{tablenotes}
	\item The aggregate coverage (Agg.), differential coverage (Diff.) and overlap (Over.) metrics were calculated for random and segregated sampling on a 256$\times$256 grid. Differences in each metric between segregated versus random patterns are listed for N = 2, 3, 4, 6, 8, 10 and R = 2, 4, 6, 8.   
\end{tablenotes}
\end{threeparttable}
\end{table}

Multi-contrast T$_2$-weighted images of the brain phantom were simulated based on Eq.~\ref{eq:se}. Relaxation parameters were identical to those used in bSSFP simulations. The parameters of SE were $\alpha = 90^o–180^o$, TR = 2800 ms, and TE = (60, 100, 140) ms corresponding to N = 3. PE reconstructions were computed on acquisitions undersampled with R = 3 in two phase-encode dimensions using random, low-correlation and segregated ($\mu = 0$) sampling.

The interaction between noise level and sampling performance was examined on bSSFP and T$_2$-weighted images. Bivariate Gaussian noise with zero mean and variance ranging from 10$^{-6}$ to 10$^{-2}$ was added. PE reconstructions were performed on noisy data undersampled with random and segregated ($\mu = 0$) patterns. The following parameters were used: N = (4, 6, 8) and R = N for bSSFP images, and N = 3 and R = 3 for T$_2$-weighted images.  

To examine the effect of k-space coverage on image quality, the parameter $\mu$ in Eq.~\ref{eq:dmod1} was tuned to systematically vary aggregate coverage from that of random sampling to that of segregated sampling. Phase-cycled bSSFP images of the brain were undersampled for $\mu$ = (0, 0.2, 0.4, 0.6, 0.8, 1.0). At each value of $\mu$, PE reconstructions were performed for N = 4, 6, 8 and R = N. The simulations were repeated for 10 independent sets of sampling patterns. 

Reconstruction quality was evaluated via comparisons to Fourier reconstructions of fully-sampled acquisitions. For bSSFP images, comparisons were performed on the combination image across phase-cycles. For multi-contrast images, comparisons were performed individually on each contrast image. Mean-squared error (MSE) maps, peak signal-to-noise ratio (PSNR), and structural similarity index (SSIM) were measured. Separate reconstructions were obtained for 10 different cross sections with different instances of sampling patterns. Significance was assessed with Wilcoxon signed-rank tests.

\begin{figure}[t]
  \begin{center}
    \includegraphics[width = \swidth]{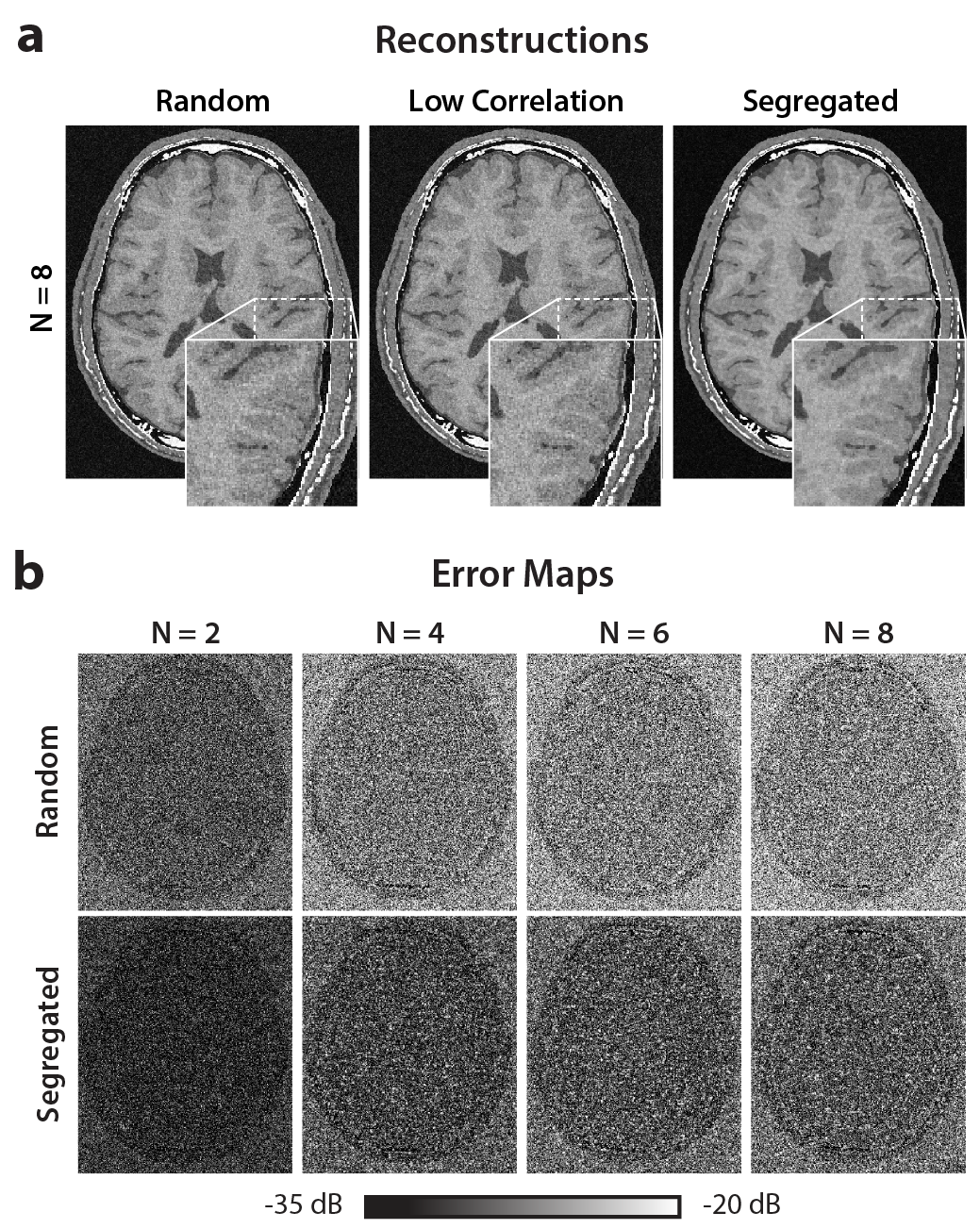}
    \caption{Multiple T$_1$-weighted phantom images were obtained by undersampling the same data by N separate patterns. Zero-filled Fourier (ZF) reconstructions were summed across acquisitions. \textbf{(a)} Images with random and segregated sampling at N = 8. Zoomed-in portions are shown in small display windows. Segregated sampling substantially reduces aliasing interference. \textbf{(b)} Error between ZF reconstructions and a fully-sampled reference image is shown in logarithmic scale (see colorbar) for N = [2 8]. At all N, segregated sampling reduces reconstruction error across the FOV compared to random sampling.
    }
    \label{fig:phantt1}
  \end{center}
\end{figure}

\begin{figure}[t]
  \begin{center}
   \includegraphics[width = \swidth]{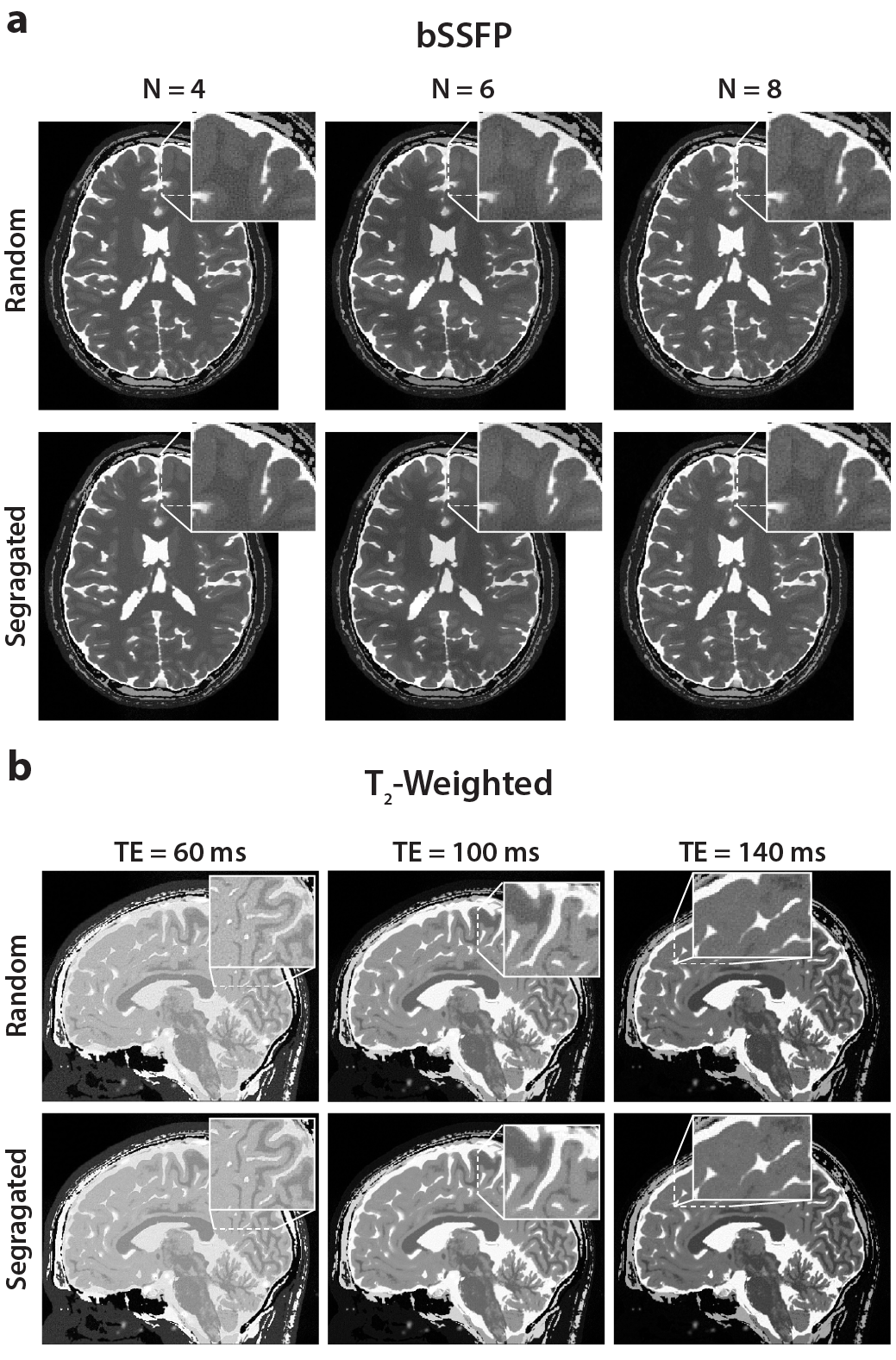}
    \caption{Brain phantom acquisitions were undersampled with random and segregated sampling, and profile-encoding reconstructions were performed. \textbf{(a)} Phase-cycled bSSFP images for N = 4, 6, 8 and R = N. \textbf{(b)} T$_2$-weighted images for TE = 60, 100 and 140 ms and R = 3. For both bSSFP and T$_2$-weighted images, reconstructions following segregated sampling have reduced interference from residual aliasing and noise compared to random sampling.
    }
    \label{fig:phantc}
  \end{center}
\end{figure}

\subsection*{Experiments}
The sampling strategies were demonstrated in vivo using a 3 T Siemens scanner (maximum gradient strength of 45 mT/m and slew rate of 200 T/m/s) and a 32-channel coil. First, brain images were collected using a three-dimensional (3D) bSSFP sequence with $\alpha = 30^o$, TR/TE = 5.1/2.65 ms, field-of-view (FOV) = 22 cm, 0.85-mm isotropic resolution, A/P and R/L phase-encoding, N = 8 with $\Delta\phi$ equisapced in the range [0 2$\pi$). Second, T$_2$-weighted brain images were collected using a 3D turbo spin-echo (TSE) sequence with $\alpha = 90-170^o$, TR = 3000 ms, TE = 145, 257 and 320 ms, FOV = 25.6 cm, 1-mm isotropic resolution, A/P and R/L phase-encoding. Each acquisition was linearly combined across coils to obtain a single-channel dataset \cite{Ilicak:FJpKoYYb}. Protocols were approved by the local ethics committee, and informed consent was obtained. 

In vivo acquisitions were undersampled in two phase-encode dimensions using random, low-correlation, and segregated ($\mu = 0$) sampling. For bSSFP datasets, N = 2, 4, 6 and 8, and R = N were used. For T$_2$-weighted datasets, N = 3 and R = 3 were used. The effect of aggregate k-space coverage on sampling performance was assessed on bSSFP images for N = 4, 6, 8 and R = N, while $\mu$ was set as (0, 0.2, 0.4, 0.6, 0.8, 1.0). 

PE reconstructions were performed on undersampled data, and reconstruction quality was assessed by PSNR and SSIM between the reconstructed images and reference images obtained via fully-sampled Fourier reconstruction. This procedure was repeated across 10 different cross sections with different instances of sampling patterns. Significance was assessed with Wilcoxon signed-rank tests.

\begin{table}[t]
	\scriptsize
	\setlength{\tabcolsep}{3.6pt}
	\centering
	\caption{Reconstructions of Phantom Images}
	\label{tab:phant} 
	\begin{threeparttable}
	\begin{tabular*}{\sswidth}{@{\extracolsep{\fill}}l|l|cccc} 
		\multicolumn{6}{c}{} \\[0.25ex] 
\multicolumn{6}{c}{\textbf{T$_1$-weighted Images}} \\[.5ex]
\hline  
\multicolumn{2}{c|}{\T \textbf{}} & \textbf{N = 2} & \textbf{N = 4} & \textbf{N = 6} & \B \textbf{N = 8} \\
\hline
\hline \T
\multirow{2}{*}{\textbf{Random}}    & PSNR & 26.9$\pm$0.1 & 23.9$\pm$0.1 & 23.0$\pm$0.1 & 22.6$\pm$0.1 \\ 
                                    & SSIM & 52.5$\pm$0.7 & 43.7$\pm$0.6 & 41.4$\pm$0.6 & 40.1$\pm$0.6 \\ \hline \T
\multirow{2}{*}{\textbf{Low Corr}}  & PSNR & 27.0$\pm$0.1 & 23.9$\pm$0.1 & 23.1$\pm$0.1 & 22.6$\pm$0.1 \\
                                    & SSIM & 52.7$\pm$0.7 & 43.9$\pm$0.7 & 41.5$\pm$0.5 & 40.1$\pm$0.6 \\ \hline \T
\multirow{2}{*}{\textbf{Segregated}}& PSNR & 30.5$\pm$0.1 & 27.9$\pm$0.1 & 26.8$\pm$0.1 &  26.2$\pm$0.1 \\
                                    & SSIM & 63.6$\pm$0.8 & 57.0$\pm$0.8 & 53.8$\pm$0.6 &  52.0$\pm$0.6 \\ \hline
\end{tabular*}

\begin{tabular*}{\sswidth}{@{\extracolsep{\fill}}l|l|cccc} 
		\multicolumn{6}{c}{} \\[0.25ex] 
\multicolumn{6}{c}{\textbf{bSSFP Images}} \\[.5ex]
\hline  
\multicolumn{2}{c|}{\T \textbf{}} & \textbf{N = 2} & \textbf{N = 4} & \textbf{N = 6} & \B \textbf{N = 8} \\
\hline
\hline \T
\multirow{2}{*}{\textbf{Random}}    & PSNR & 35.7$\pm$0.1 & 32.2$\pm$0.2 & 31.2$\pm$0.1 & 30.0$\pm$0.2 \\ 
                                    & SSIM & 91.0$\pm$0.2 & 81.8$\pm$0.4 & 77.9$\pm$0.3 & 72.8$\pm$0.5 \\ \hline \T
\multirow{2}{*}{\textbf{Low Corr}}  & PSNR & 35.7$\pm$0.1 & 32.2$\pm$0.3 & 31.2$\pm$0.2 & 30.0$\pm$0.2 \\
                                    & SSIM & 91.0$\pm$0.2 & 81.8$\pm$0.7 & 77.7$\pm$0.6 & 72.2$\pm$0.5 \\ \hline \T
\multirow{2}{*}{\textbf{Segregated}}& PSNR & 35.7$\pm$0.1 & 32.7$\pm$0.3 & 32.1$\pm$0.1 & 30.4$\pm$0.2 \\
                                    & SSIM & 91.1$\pm$0.2 & 82.8$\pm$0.8 & 79.7$\pm$0.5 & 73.9$\pm$0.4 \\ \hline
\end{tabular*}

\begin{tabular*}{\sswidth}{@{\extracolsep{\fill}}l|l|ccc} 
		\multicolumn{5}{c}{} \\[0.25ex] 
\multicolumn{5}{c}{\textbf{T$_2$-weighted Images}} \\[.5ex]
\hline  
\multicolumn{2}{c|}{\T \textbf{}} & \textbf{TE=60ms} & \textbf{TE=100ms} & \textbf{TE=140ms} \\
\hline
\hline \T
\multirow{2}{*}{\textbf{Random}} & PSNR & 28.6$\pm$1.0 & 30.2$\pm$0.5 & 28.8$\pm$0.5\\ 
& SSIM & 77.2$\pm$1.1 & 84.2$\pm$0.7 & 86.5$\pm$0.4\\ 
\hline \T
\multirow{2}{*}{\textbf{Low Corr}}   & 
PSNR & 28.7$\pm$1.0 & 30.2$\pm$0.5 & 28.7$\pm$0.5\\
                            & 
SSIM  & 77.3$\pm$1.2 & 84.2$\pm$0.7 & 86.5$\pm$0.4\\ \hline \T
\multirow{2}{*}{\textbf{Segregated}}  & PSNR     & 30.5$\pm$1.2 & 33.0$\pm$0.5 & 30.3$\pm$0.4 \\
                            & SSIM & 83.0$\pm$0.8 & 89.6$\pm$0.5 & 89.7$\pm$0.2\\ \hline
\end{tabular*}

\begin{tablenotes}
	\item PSNR (dB) and SSIM (\%) measurements on T$_1$-weighted (top), bSSFP (middle) and T$_2$-weighted (bottom) images of the brain phantom. Random, low-correlation and segregated sampling were performed at N = [2 8] for T$_1$-weighted and bSSFP images, and at N = 3 (three echo times) for T$_2$-weighted images. Metrics are reported as mean$\pm$std across 10 different cross sections.
\end{tablenotes}
\end{threeparttable}
\end{table}

\begin{figure}[t]
  \begin{center} 
    \includegraphics[width = \swidth]{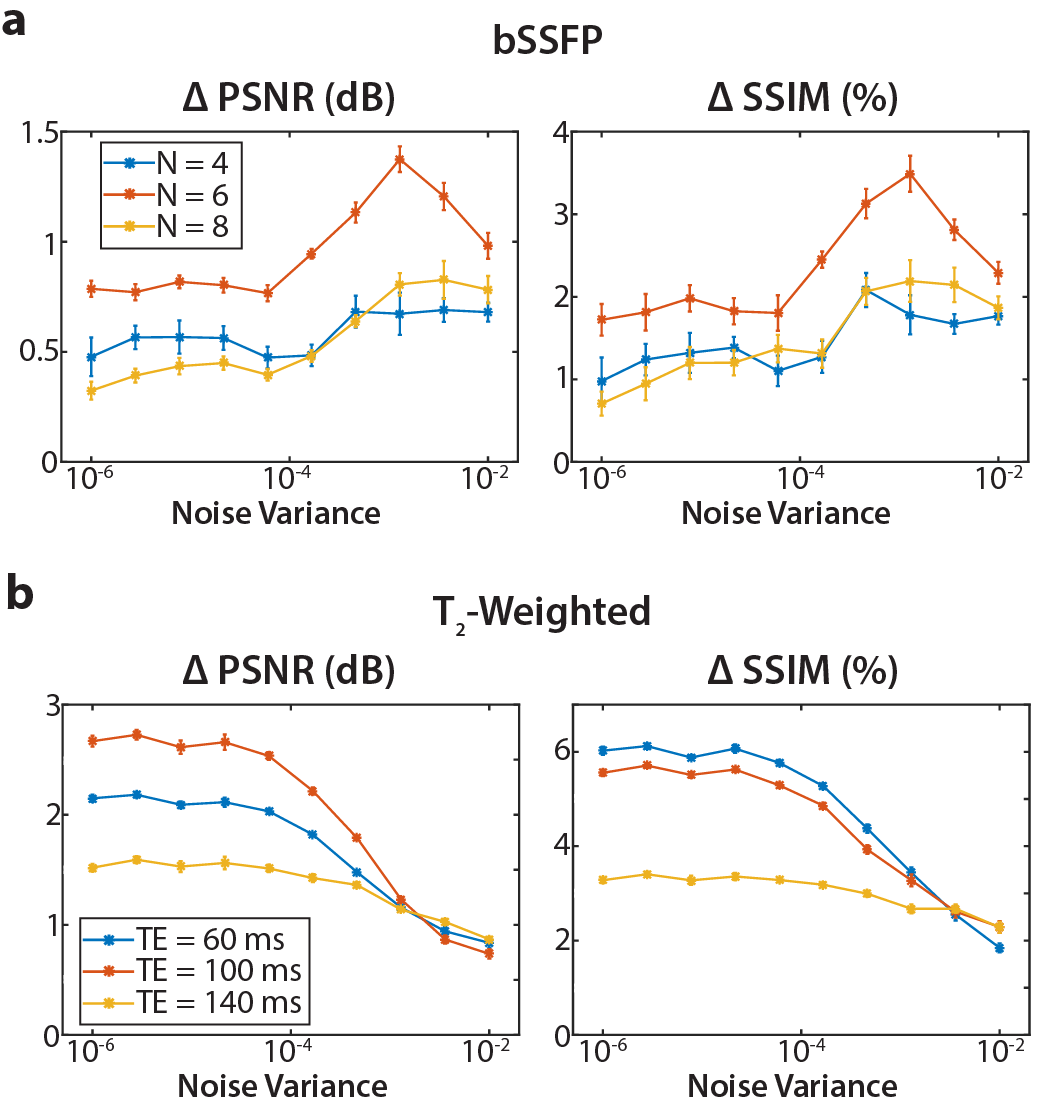}
    \caption{The effect of noise level on sampling performance was examined on phase-cycled bSSFP and T$_2$-weighted brain phantom images. Bivariate Gaussian noise with zero mean and variance in [10$^{-6}$ 10$^{-2}$] was added, and PE reconstructions were performed on noisy data undersampled with random and segregated patterns (N = R). Improvements in PSNR and SSIM with segregated sampling over random sampling were calculated. Error bars display mean$\pm$std across 10 independent sets of noise instance and sampling patterns. \textbf{(a)} Improvements in bSSFP images for N = (4, 6, 8). \textbf{(b)} Improvements in T$_2$-weighted images for three echo times (TE). Segregated sampling achieves superior performance for a broad range noise levels. 
    }
    \label{fig:phantnoise}
  \end{center}
\end{figure}

\begin{figure}[t]
  \begin{center} 
    \includegraphics[width = \swidth]{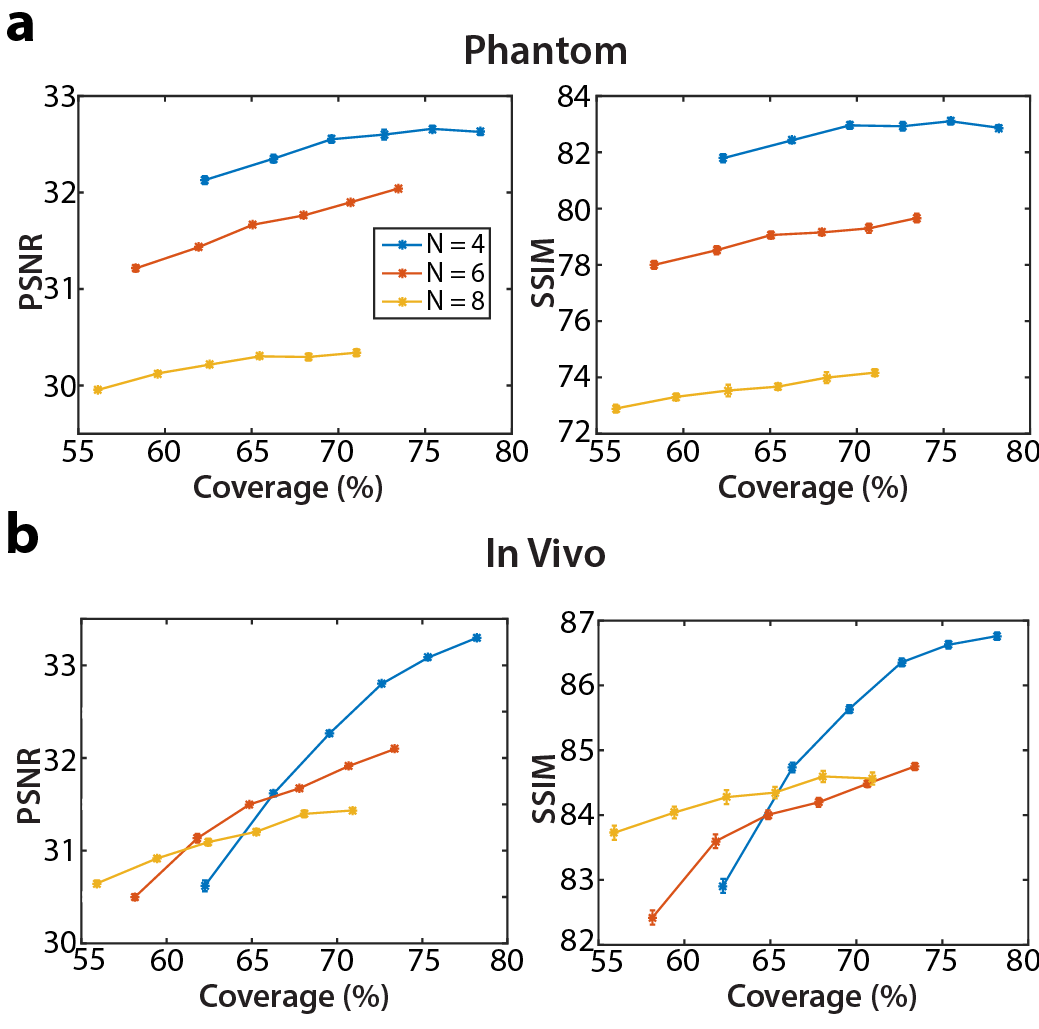}
    \caption{The parameter $\mu$ in Eq.~\ref{eq:dmod1} was tuned to systematically vary aggregate k-space coverage from that of random sampling to that of segregated sampling. Balanced SSFP datasets were undersampled for $\mu$ = (0, 0.2, 0.4, 0.6, 0.8, 1.0). PE reconstructions were obtained for N = 4, 6, 8 and R = N. Error bars display mean$\pm$std across 10 independent sets of sampling patterns. \textbf{(a)} PSNR and SSIM for brain phantom images. \textbf{(b)} PSNR and SSIM for in vivo brain images. For both datasets and at all N, reconstruction quality improves persistently as the aggregate coverage is broadened. 
    }
    \label{fig:imstats}
  \end{center}
\end{figure}

\begin{figure}[t]
  \begin{center}
    \includegraphics[width = \swidth]{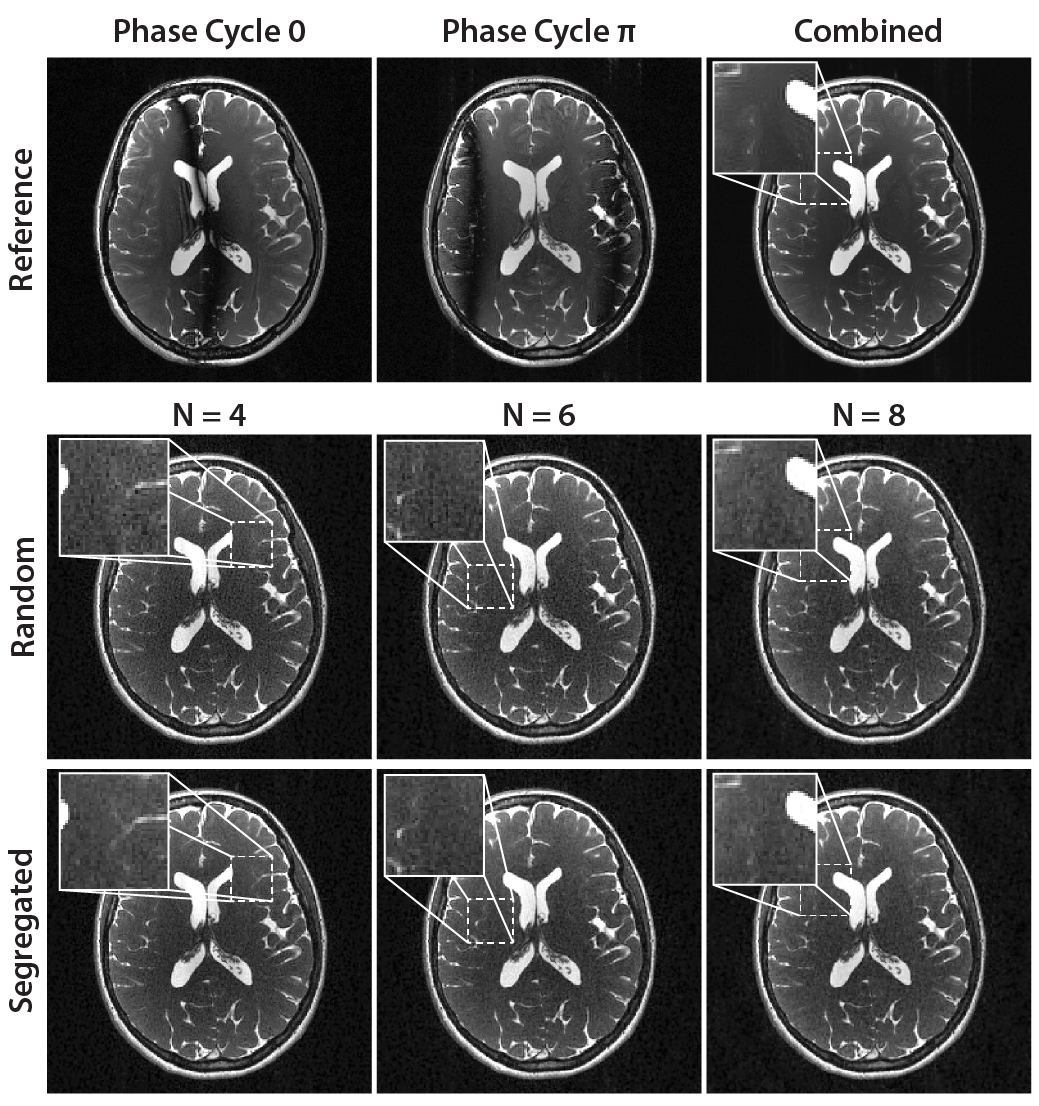}
    \caption{In vivo bSSFP acquisitions of the brain were undersampled with random and segregated sampling. PE reconstructions were performed for N = 4, 6, 8 and R = N. Segregated sampling provides reduced reconstruction error compared to random sampling. Zoomed-in display windows show detailed features that are poorly depicted with random sampling. These features are sensitively recovered with segregated sampling. 
    }
    \label{fig:vivossfp}
  \end{center}
\end{figure}

\section{Results}

\subsection*{Simulation Analyses}
Sampling methods were first compared in terms of their aggregate coverage, differential coverage and overlap. Representative patterns from random, low-correlation, and segregated sampling are shown in Fig. \ref{fig:masks}. Low-correlation sampling yields limited improvement over random sampling, thus subsequent comparisons focused on segregated versus random methods. Measurements for N = [2 16] and R = [2 8] are plotted in \Cref{fig:patcomp} and summarized in \Cref{tab:patcomp}. While random sampling can leave 10-30\% of k-space uncovered even at N = 16, segregated sampling achieves full coverage within N = 2$\times$R acquisitions. This increased coverage is accompanied by expanded differential coverage within individual patterns and reduced overlap across patterns, particularly for higher R and lower N. Segregated sampling achieves 14.6$\pm$1.4\% (mean$\pm$std across N, where R=N) higher aggregate coverage, 26.0$\pm$0.9\% higher differential coverage, and 20.2$\pm$3.8\% reduced overlap relative to random sampling.

To demonstrate segregated sampling, repeated T$_1$-weighted acquisitions of a brain phantom were first simulated. ZF reconstructions are shown in \Cref{fig:phantt1}, and PSNR and SSIM measurements are listed in \Cref{tab:phant} for N = [2 8]. Segregated sampling reduces aliasing interference compared to alternative methods, with 3.8$\pm$0.2 dB higher PSNR and 12.2$\pm$0.9\% higher SSIM than random sampling (p$<$0.05).

Following this demonstration on a linear reconstruction, segregated sampling was evaluated based on PE reconstructions of bSSFP acquisitions. Reconstructions  of the brain phantom are shown in \Cref{fig:phantc}a, and PSNR and SSIM on combined bSSFP images are listed in \Cref{tab:phant} for N = [2 8]. Segregated sampling achieves 0.5$\pm$0.3 dB higher PSNR and 1.0$\pm$0.7\% higher SSIM compared to random sampling (p$<$0.05, except N=2 where they perform similarly). Next, multi-contrast T$_2$-weighted acquisitions of the brain phantom were examined. Reconstructions at three echo times (TE) are shown in \Cref{fig:phantc}b, and PSNR and SSIM for N = 3 are listed in \Cref{tab:phant}. Segregated sampling improves PSNR by 2.1$\pm$0.7 dB (mean$\pm$std across TE) and SSIM by 4.8$\pm$1.4\% over random sampling (p$<$0.05). 

To assess reliability against noise, reconstruction quality was evaluated across a broad range of noise levels. \Cref{fig:phantnoise} displays the difference in PSNR and SSIM between segregated and random sampling. For both bSSFP and T$_2$-weighted images, segregated sampling improves image quality across the entire noise range. The improvements grow for higher noise levels in bSSFP images, where quality metrics were calculated on combined images that average data across acquisitions. In contrast, metrics were calculated on each T$_2$-weighted image without averaging, thus the relatively larger improvements in this case decline with higher noise. 

Lastly, the effect of aggregate coverage on sampling performance was examined. Phase-cycled bSSFP acquisitions of the brain phantom were undersampled for varying values of $\mu$, which controls the aggregate coverage. PSNR and SSIM are plotted as a function of aggregate coverage in \Cref{fig:imstats}a. At all N, PSNR and SSIM improve consistently with increased aggregate coverage. Taken together, these results suggest that segregated sampling captures greater information about tissue structure and leads to improved CS recovery due to its expanded coverage and reduced pattern overlap. 

\subsection*{In Vivo Analyses}
The proposed strategy was demonstrated on in vivo bSSFP images of the brain. Random, low-correlation and segregated sampling were compared in terms of the respective PE reconstructions combined across acquisitions. Representative reconstructions for N = [4 8] are shown in \Cref{fig:vivossfp}. Segregated sampling reduces residual errors in bSSFP images. Furthermore, some detailed features that are poorly depicted with random sampling are sensitively recovered with segregated sampling. These observations are supported by PSNR and SSIM measurements listed in \Cref{tab:vivo}. Segregated sampling yields 1.8$\pm$0.8 dB (mean$\pm$std across N) higher PSNR and 2.5$\pm$1.3\% higher SSIM than random sampling (p$<$0.05). 

To examine the effect of $\mu$ on sampling performance, phase-cycled bSSFP acquisitions of the brain were reconstructed with PE for varying values of $\mu$. PSNR and SSIM measurements are plotted in \Cref{fig:imstats}b. Similar to results obtained from phantom simulations, expanding aggregate coverage steadily improves quality of in vivo image reconstructions at all number of acquisitions. 

\begin{figure}[h]
  \begin{center} 
    \includegraphics[width = \swidth]{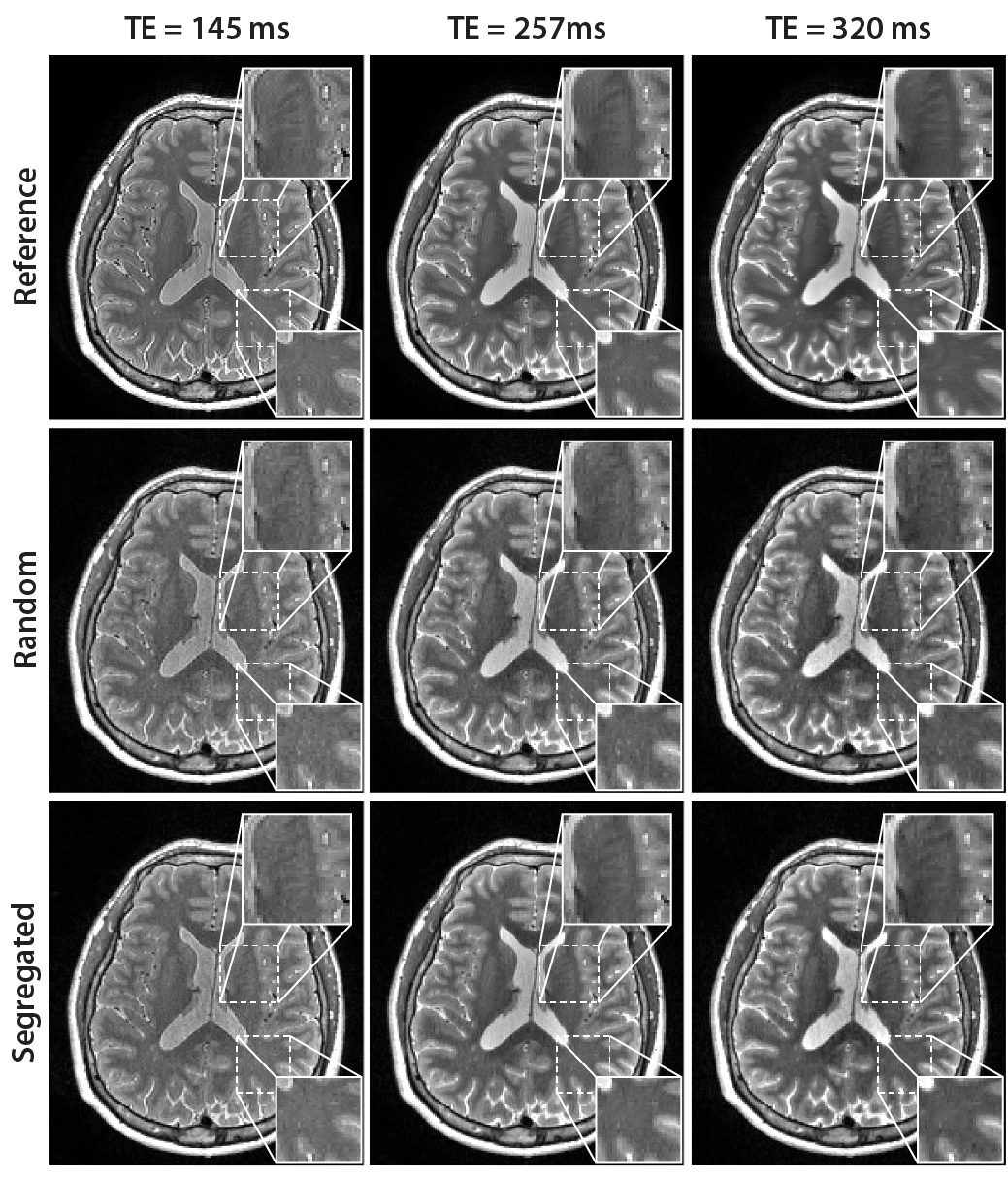}
    \caption{In vivo T$_2$-weighted acquisitions (at TE = 145, 257 and 320 ms) of the brain were undersampled with random and segregated sampling. \textbf{(a)} PE reconstructions were performed for N = 3 and R = N. Segregated sampling enables improved reconstructions due to reduced aliasing and noise interference. Zoomed-in display windows show limited-contrast or small features that are relatively more visible with segregated sampling. 
    }
    \label{fig:vivot2}
  \end{center}
\end{figure}

Next, in vivo multi-contrast T$_2$-weighted images of the brain were considered. Reconstructions at N = 3 displayed in \Cref{fig:vivot2} demonstrate improved quality with segregated sampling. Several limited-contrast or small features are relatively more visible with segregated sampling. Respective PSNR and SSIM measurements are listed in \Cref{tab:vivo}. Segregated sampling yields 0.6$\pm$0.1 dB (mean$\pm$std across TE) higher PSNR (p$<$0.05) and 1.7$\pm$1.4\% higher SSIM than random sampling (p$<$0.05, except TE = 320 ms where they perform similarly).

\section{Discussion}
Multiple acquisitions are typically collected to enhance quality or information content of MR images. Prolonged scan times can then be avoided through sparse recovery of undersampled acquisitions. To improve sparse recovery, here we proposed a segregated sampling method that statistically minimizes overlap across multiple-acquistion patterns to extend aggregate k-space coverage.

Several previous reports considered sampling strategies for multiple-acquisition data. Earlier work focused on generating incoherent patterns across separate acquisitions \cite{Lustig:2007cu}. For this purpose, each individual acquisition was accelerated via a distinct random pattern drawn from a common sampling density \cite{Doneva:2010it,Bilgic:2011jv,Huang:2014ca}. Although random sampling theoretically promises successful CS recovery, naive random selection can generate gaps or clusters across the acquisition dimension \cite{Lustig:2010hs}. In turn, a k-space gap can impair the recovery of unacquired data, whereas a k-space cluster can reduce scan efficiency by collecting redundant information. 

To prevent gaps or clusters across temporal frames, recent studies on dynamic MRI incorporated deterministic criteria for sample selection \cite{Ahmad:2015iz, Kim:2015fu, Levine:2016jt}. With similar motivations, we recently proposed low-correlation sampling to reduce pattern overlap in bSSFP imaging \cite{Cukur:2015ic}. While low-correlation sampling reduces aliasing artifacts, it uses a search procedure following pattern generation that is suboptimal for minimizing overlap. In contrast, segregated sampling reduces pattern overlap during pattern generation, yielding greater coverage.

Here the enhanced performance of segregated sampling was demonstrated for phase-cycled bSSFP and multi-contrast imaging. The quality improvement in bSSFP images is relatively higher for in vivo datasets compared to simulations (without noise), and the reverse is observed for multi-contrast images. Our analyses also indicate that, at higher noise levels typically encountered in practice, improvements in combined bSSFP images increase whereas those in individual T$_2$-weighted images decrease. Therefore, the observed differences between simulations and in vivo experiments might be attributed to varying noise levels. The precise level of improvement will depend on sequence parameters, tissue structure and experimental conditions. Regardless, segregated sampling is expected to outperform random sampling without imposing any additional computational burden.  

A number of avenues can be explored to further improve segregated sampling. Currently, the proposed method optimizes parameters of the polynomial sampling density based on the PSF of the resultant patterns. However, the spectra of MRI images are not guaranteed to strictly follow a power law, and thus a polynomial density may be suboptimal. Previous work suggests that a template of the power spectrum can be used to effectively capture the energy in MRI data \cite{Knoll:2011be,Liang:2012cq,Raja:2014gg}. Similarly, the initial sampling density in segregated sampling could be designed to match the spectrum estimates for particular anatomies and MRI contrasts.

Another improvement concerns pattern generation based on the chosen sampling density. The current method is stochastic and could result in spatial-frequency gaps or clusters within individual patterns. If excessive, gap/cluster formation may be problematic for multi-coil imaging. In such cases, the Poisson-disc algorithm could be leveraged to enforce more uniform sampling within each pattern \cite{Lustig:2010hs}.   

Segregated sampling generates individual patterns sequentially. It is possible that later patterns in the sequence are more constrained in terms of the k-space distribution of sample locations. In this study, we did not observe any degradation in the point spread function of the sampling patterns up to N = 8. Yet, potential degradations that can arise for larger N or smaller pattern sizes might be alleviated by implementing more conservative changes in sampling density across acquisitions. Alternatively, k-space can be split into annular segments \cite{Cukur:2008ir}, and the proposed method can be performed on each segment separately. To achieve more balanced sampling, the pattern-generation order for N acquisitions might be randomized across segments.  
 
The current study has several limitations that can be addressed in future work. First, segregated sampling was primarily demonstrated for acceleration in the two-phase encode dimensions of 3D Cartesian acquisitions. Note, however, that adaptation of the proposed method to 1D acceleration in 2D acquisitions is rather straightforward. Second, segregated sampling was demonstrated on coil-combined data here to simplify reconstructions by factoring out the coil dimension. It is expected that the demonstrated results will carry over to coil-by-coil reconstructions \cite{Cukur:2010ey} and joint reconstructions across coils and acquisitions. However, the relative levels of improvement provided by segregated sampling and by the encoding information from coil arrays remain to be investigated.

To conclude, segregated sampling improves multiple-acquisition MRI reconstructions by achieving incoherent aliasing both within and across acquisitions. Here we demonstrated improvements for phase-cycled bSSFP and multiple-contrast data compared to conventional sampling methods. In principle, the proposed method can also benefit other applications where multiple acquisitions are critical such as peripheral angiography \cite{Kwon:2014cy} and fat/water separation \cite{ATRDIXON,Quist:2012kx}.


\begin{table}
	\scriptsize
	\setlength{\tabcolsep}{3.6pt}
	\centering
	\caption{Reconstructions of In Vivo Images}
	\label{tab:vivo} 
	\begin{threeparttable}
	\begin{tabular*}{\sswidth}{@{\extracolsep{\fill}}l|l|cccc} 
		\multicolumn{6}{c}{} \\[0.25ex] 
\multicolumn{6}{c}{\textbf{bSSFP Images}} \\[.5ex] \hline  
\multicolumn{2}{c|}{\T \textbf{}} & \textbf{N = 2} & \textbf{N = 4} & \textbf{N = 6} & \textbf{N = 8} \\ \hline \hline \T
\multirow{2}{*}{\textbf{Random}}    & PSNR & 33.9$\pm$0.7 & 30.4$\pm$0.6 & 30.3$\pm$0.7 & 30.4$\pm$0.5 \\ 
                                    & SSIM & 88.4$\pm$1.3 & 81.6$\pm$1.5 & 81.1$\pm$2.2 & 82.3$\pm$1.5 \\ \hline \T
\multirow{2}{*}{\textbf{Low Corr}}  & PSNR & 33.9$\pm$0.6 & 30.5$\pm$0.4 & 30.5$\pm$0.6 & 30.3$\pm$0.5 \\
                                    & SSIM & 88.2$\pm$1.2 & 81.8$\pm$1.5 & 81.5$\pm$1.8 & 81.9$\pm$1.5 \\ \hline \T
\multirow{2}{*}{\textbf{Segregated}}& PSNR & 36.0$\pm$0.6 & 33.2$\pm$0.4 & 32.0$\pm$0.5 & 31.3$\pm$0.4 \\
                                    & SSIM & 90.9$\pm$0.9 & 85.8$\pm$1.2 & 83.6$\pm$1.5 & 83.3$\pm$1.3 \\ \hline
\end{tabular*}

\begin{tabular*}{\sswidth}{@{\extracolsep{\fill}}l|l|ccc} 
		\multicolumn{5}{c}{} \\[0.25ex] 
\multicolumn{5}{c}{\textbf{T$_2$-weighted Images}} \\[.5ex] \hline  
\multicolumn{2}{c|}{\T \textbf{}} & \textbf{TE=145ms} & \textbf{TE=257ms} & \textbf{TE=320ms} \\ \hline \hline \T
\multirow{2}{*}{\textbf{Random}}    & PSNR & 28.1$\pm$0.2 & 29.7$\pm$0.3 & 29.6$\pm$0.3 \\ 
                                    & SSIM & 80.3$\pm$0.6 & 80.7$\pm$0.7 & 81.6$\pm$0.5 \\ \hline \T
\multirow{2}{*}{\textbf{Low Corr}}  & PSNR & 28.2$\pm$0.2 & 29.7$\pm$0.3 & 29.6$\pm$0.3 \\
                                    & SSIM & 81.0$\pm$0.7 & 80.2$\pm$0.3 & 81.6$\pm$0.5 \\ \hline \T
\multirow{2}{*}{\textbf{Segregated}}& PSNR & 28.6$\pm$0.3 & 30.5$\pm$0.4 & 30.1$\pm$0.3 \\
                                    & SSIM & 82.6$\pm$0.9 & 83.3$\pm$0.5 & 81.7$\pm$0.7 \\ \hline
\end{tabular*}

\begin{tablenotes}
	\item PSNR and SSIM measurements on in vivo bSSFP (upper) and T$_2$-weighted (lower) images of the brain. Random, low-correlation and segregated sampling was performed at N = 2, 4, 6, 8 for bSSFP images, and at N = 3 (three echo times) for T$_2$-weighted images. Metrics are reported as mean$\pm$std across 10 different cross sections.
\end{tablenotes}
\end{threeparttable}
\end{table}


\bibliographystyle{IEEEtran}

\end{document}